# Quantum Associative Memory

Dan Ventura and Tony Martinez

Neural Networks and Machine Learning Laboratory (*http://axon.cs.byu.edu*)

Department of Computer Science

Brigham Young University

*dan@axon.cs.byu.edu*, *martinez@cs.byu.edu*

**Abstract**  This paper combines quantum computation with classical neural network theory to produce a quantum computational learning algorithm. Quantum computation uses microscopic quantum level effects to perform computational tasks and has produced results that in some cases are exponentially faster than their classical counterparts. The unique characteristics of quantum theory may also be used to create a quantum associative memory with a capacity exponential in the number of neurons. This paper combines two quantum computational algorithms to produce such a quantum associative memory. The result is an exponential increase in the capacity of the memory when compared to traditional associative memories such as the Hopfield network. The paper covers necessary high-level quantum mechanical and quantum computational ideas and introduces a quantum associative memory. Theoretical analysis proves the utility of the memory, and it is noted that a small version should be physically realizable in the near future.

## 1. Introduction

The field of neural networks seeks, among other things, to develop algorithms for imitating in some sense the functionality of the brain. One particular area of interest is that of associative pattern recall. The field of quantum computation (QC) investigates the power of the unique characteristics of quantum systems used as computational machines. This paper combines results from both of these fields to produce a new quantum computational learning algorithm. This contributes significantly to both the field of quantum computation and to the field of neural networks. The field of neural networks benefits by the introduction of a quantum associative memory with a storage capacity exponential in the number of neurons. The contribution to QC is in the form of a new quantum algorithm capable of results that appear to be impossible using classical computational methods.

Assume a set $\wp$ of $m$ binary patterns of length $n$. We consider the problem of associative pattern completion -- learning to produce one of the full patterns when presented with only a partial pattern. The trivial solution is simply to store the set of patterns as a lookup table or RAM. There are two reasons why this is not always the best solution. First, it requires that a unique address be associated with and remembered for each pattern. Second, the lookup table requires $mn$ bits in order to store all the patterns. It is often desirable to be able to recall the patterns in an associative fashion, thus eliminating the need for explicit addressing. That is, given a partial pattern one would like to be able to "fill in" a reasonable guess as to the rest of the pattern. This may also be considered a form of generalization as the partial pattern may never have been seen during the learning of the pattern set $\wp$. Further, it would of course be beneficial if a smaller representation was possible.

To this end, various classical associative memory schemes have been proposed, perhaps the most well known being the Hopfield network [Hop82] and the bidirectional associative memory (BAM) [Kos88]. These neural approaches to the pattern completion problem allow for associative pattern recall, but suffer severe storage restrictions. Storing patterns of length $n$ requires a network of $n$ neurons, and the number of patterns, $m$, is then limited by $m \leq kn$, where typically $.15 \leq k \leq .5$. This paper offers improvement by proposing a quantum associative memory that maintains the ability to recall patterns associatively while offering a storage capacity of $O(2^n)$ using only $n$ neurons.

The field of quantum computation, which applies ideas from quantum mechanics to the study of computation, was introduced in the mid 1980's [Ben82] [Deu85] [Fey86]. For a readable introduction to quantum computation see [Bar96]. The field is still in its infancy and very theoretical but offers exciting possibilities for the field of computer science -- perhaps the most notable to date being the discovery of quantum computational algorithms for computing discrete logarithms and prime factorization in polynomial time, two problems for which no known classical polynomial time solutions exist [Sho97]. These algorithms provide theoretical proof not only that interesting computation can be performed at the quantum level but also that it may in some cases have distinct advantages over its classical cousin. Very recently several groups have produced exciting experimental results by successfully implementing quantum algorithms on small-scale nuclear magnetic resonance (NMR) quantum computers (see for example [Jon98] and [Chu98]).

Artificial neural networks (ANNs) seek to provide ways for classical computers to learn rather than to be programmed. As quantum computer technology continues to develop, artificial neural network methods that are amenable to and take advantage of quantum mechanical properties will become possible. In particular, can quantum mechanical properties be applied to ANNs for problems such as associative memory? Recently, work has been done in the area of combining classical artificial associative memory with ideas from the field of quantum mechanics. Perus details several interesting mathematical analogies between quantum theory and neural network theory [Per96] [Beh96]. [Ven98a] goes a step further by proposing an actual model for a quantum associative memory and [Ven98c] further develops this model by exhibiting a physically realizable quantum system for acting as an associative memory. The work here extends the work introduced in [Ven98c], by further developing the ideas, presenting examples and providing rigorous theoretical analysis.

This paper presents a unique reformulation of the pattern completion problem into the language of wave functions and operators. This reformulation may be generalized to a large class of computational learning problems, opening up the possibility of employing the capabilities of quantum computational systems for the solution of computational learning problems. Section 2 presents some basic ideas from quantum mechanics and introduces quantum computation and some of its early successes. Since neither of these subjects can be properly covered here, references for further study are provided. Section 3 discusses in some detail two quantum algorithms, one for storing a set of patterns in a quantum system and one for quantum search. The quantum associative memory that is the main result of this paper is presented in section 4 along with theoretical analysis of the model, and the paper concludes with final remarks and directions for further research in section 5.

## 2. Quantum Computation

Quantum computation is based upon physical principles from the theory of quantum mechanics (QM), which in many ways is counterintuitive. Yet it has provided us with perhaps the most accurate physical theory (in terms of predicting experimental results) ever devised by science. The theory is well-established and is covered in its basic form by many textbooks (see for example [Fey65]). Several necessary ideas that form the basis for the study of quantum computation are briefly reviewed here.

### 2.1. Linear Superposition

*Linear superposition* is closely related to the familiar mathematical principle of linear combination of vectors. Quantum systems are described by a wave function $\psi$ that exists in a Hilbert space [You88]. The Hilbert space has a set of states, $|\phi_i\rangle$, that form a basis, and the system is described by a quantum state,

$$|\psi\rangle = \sum_i c_i |\phi_i\rangle. \qquad (1)$$

$|\psi\rangle$ is said to be in a linear superposition of the basis states $|\phi_i\rangle$, and in the general case, the coefficients $c_i$ may be complex. Use is made here of the Dirac bracket notation, where the ket $|\cdot\rangle$ is analogous to a column vector, and the bra $\langle\cdot|$ is analogous to the complex conjugate transpose of the ket. In quantum mechanics the Hilbert space and its basis have a physical interpretation, and this leads directly to perhaps the most counterintuitive aspect of the theory. The counter intuition is this -- at the microscopic or quantum level, the state of the system is described by the wave function $\psi$, that is, as a linear superposition of all basis states (i.e. in some sense the system is in all basis states at once). However, at the macroscopic or classical level the system can be in only a single basis state. For example, at the quantum level an electron can be in a superposition of many different energies; however, in the classical realm this obviously cannot be.

## 2.2. Coherence and Decoherence

*Coherence* and *decoherence* are closely related to the idea of linear superposition. A quantum system is said to be coherent if it is in a linear superposition of its basis states. A result of quantum mechanics is that if a system that is in a linear superposition of states interacts in any way with its environment, the superposition is destroyed. This loss of coherence is called decoherence and is governed by the wave function $\psi$. The coefficients $c_i$ are called probability amplitudes, and $|c_i|^2$ gives the probability of $|\psi\rangle$ collapsing into state $|\phi_i\rangle$ if it decoheres. Note that the wave function $\psi$ describes a real physical system that must collapse to exactly one basis state. Therefore, the probabilities governed by the amplitudes $c_i$ must sum to unity. This necessary constraint is expressed as the unitarity condition

$$\sum_i |c_i|^2 = 1. \tag{2}$$

In the Dirac notation, the probability that a quantum state $|\psi\rangle$ will collapse into an eigenstate $|\phi_i\rangle$ is written $|\langle \phi_i | \psi \rangle|^2$ and is analogous to the dot product (projection) of two vectors. Consider, for example, a discrete physical variable called spin. The simplest spin system is a two-state system, called a spin-1/2 system, whose basis states are usually represented as $|\uparrow\rangle$ (spin up) and $|\downarrow\rangle$ (spin down). In this simple system the wave function $\psi$ is a distribution over two values (up and down) and a coherent state $|\psi\rangle$ is a linear superposition of $|\uparrow\rangle$ and $|\downarrow\rangle$. One such state might be

$$|\psi\rangle = \tfrac{2}{\sqrt{5}}|\uparrow\rangle + \tfrac{1}{\sqrt{5}}|\downarrow\rangle. \tag{3}$$

As long as the system maintains its quantum coherence it cannot be said to be either spin up or spin down. It is in some sense both at once. Classically, of course, it must be one or the other, and when this system decoheres the result is, for example, the $|\uparrow\rangle$ state with probability

$$|\langle \uparrow | \psi \rangle|^2 = \left(\tfrac{2}{\sqrt{5}}\right)^2 = .8. \tag{4}$$

A simple two-state quantum system, such as the spin-1/2 system just introduced, is used as the basic unit of quantum computation. Such a system is referred to as a quantum bit or *qubit*, and renaming the two states $|0\rangle$ and $|1\rangle$ it is easy to see why this is so.

## 2.3. Operators

*Operators* on a Hilbert space describe how one wave function is changed into another. Here they will be denoted by a capital letter with a hat, such as $\hat{A}$, and they may be represented as matrices acting on vectors. Using operators, an eigenvalue equation can be written $\hat{A}|\phi_i\rangle = a_i|\phi_i\rangle$, where $a_i$ is the eigenvalue. The solutions $|\phi_i\rangle$ to such an equation are called eigenstates and can be used to construct the basis of a Hilbert space as discussed in section 2.1. In the quantum formalism, all properties are represented as operators whose eigenstates are the basis for the Hilbert space associated with that property and whose eigenvalues are the quantum allowed values for that property. It is important to note that operators in quantum mechanics must be linear operators and further that they must be unitary so that $\hat{A}^\dagger \hat{A} = \hat{A}\hat{A}^\dagger = \hat{I}$, where $\hat{I}$ is the identity operator, and $\hat{A}^\dagger$ is the complex conjugate transpose, or adjoint, of $\hat{A}$.

## 2.4. Interference

*Interference* is a familiar wave phenomenon. Wave peaks that are in phase interfere constructively (magnify each other's amplitude) while those that are out of phase interfere destructively (decrease or eliminate each other's amplitude). This is a phenomenon common to all kinds of wave mechanics from water waves to optics. The well-known double slit experiment demonstrates empirically that at the quantum level interference also applies to the probability waves of quantum mechanics. As a simple example, suppose that the wave function described in equation (3) is represented in vector form as

$$|\psi\rangle = \frac{1}{\sqrt{5}}\begin{pmatrix} 2 \\ 1 \end{pmatrix} \qquad (5)$$

and suppose that it is operated upon by an operator $\hat{O}$ described by the following matrix,

$$\hat{O} = \frac{1}{\sqrt{2}}\begin{pmatrix} 1 & 1 \\ 1 & -1 \end{pmatrix}. \qquad (6)$$

The result is

$$\hat{O}|\psi\rangle = \frac{1}{\sqrt{2}}\begin{pmatrix} 1 & 1 \\ 1 & -1 \end{pmatrix}\frac{1}{\sqrt{5}}\begin{pmatrix} 2 \\ 1 \end{pmatrix} = \frac{1}{\sqrt{10}}\begin{pmatrix} 3 \\ 1 \end{pmatrix}, \quad (7)$$

and therefore now

$$|\psi\rangle = \frac{3}{\sqrt{10}}|\uparrow\rangle + \frac{1}{\sqrt{10}}|\downarrow\rangle. \quad (8)$$

Notice that the amplitude of the $|\uparrow\rangle$ state has increased while the amplitude of the $|\downarrow\rangle$ state has decreased. This is due to the wave function interfering with itself through the action of the operator -- the different parts of the wave function interfere constructively or destructively according to their relative phases just like any other kind of wave.

## 2.5. Entanglement

*Entanglement* is the potential for quantum states to exhibit correlations that cannot be accounted for classically. From a computational standpoint, entanglement seems intuitive enough -- it is simply the fact that correlations can exist between different qubits -- for example if one qubit is in the $|1\rangle$ state, another will be in the $|1\rangle$ state. However, from a physical standpoint, entanglement is little understood. The questions of what exactly it is and how it works are still not resolved. What makes it so powerful (and so little understood) is the fact that since quantum states exist as superpositions, these correlations somehow exist in superposition as well. When the superposition is destroyed, the proper correlation is somehow communicated between the qubits, and it is this "communication" that is the crux of entanglement. Mathematically, entanglement may be described using the density matrix formalism. The density matrix $\rho_\psi$ of a quantum state $|\psi\rangle$ is defined as

$$\rho_\psi = |\psi\rangle\langle\psi|. \quad (9)$$

For example, the quantum state

$$|\xi\rangle = \frac{1}{\sqrt{2}}|00\rangle + \frac{1}{\sqrt{2}}|01\rangle \quad (10)$$

appears in vector form as

$$|\xi\rangle = \frac{1}{\sqrt{2}}\begin{pmatrix} 1 \\ 1 \\ 0 \\ 0 \end{pmatrix}, \quad (11)$$

and it may also be represented as the density matrix

$$\rho_\xi = |\xi\rangle\langle\xi| = \frac{1}{2}\begin{pmatrix} 1 & 1 & 0 & 0 \\ 1 & 1 & 0 & 0 \\ 0 & 0 & 0 & 0 \\ 0 & 0 & 0 & 0 \end{pmatrix}, \qquad (12)$$

while the state

$$|\psi\rangle = \tfrac{1}{\sqrt{2}}|00\rangle + \tfrac{1}{\sqrt{2}}|11\rangle \qquad (13)$$

is represented as

$$\rho_\psi = |\psi\rangle\langle\psi| = \frac{1}{2}\begin{pmatrix} 1 & 0 & 0 & 1 \\ 0 & 0 & 0 & 0 \\ 0 & 0 & 0 & 0 \\ 1 & 0 & 0 & 1 \end{pmatrix} \qquad (14)$$

and the state $|\zeta\rangle = \tfrac{1}{\sqrt{3}}|00\rangle + \tfrac{1}{\sqrt{3}}|01\rangle + \tfrac{1}{\sqrt{3}}|11\rangle$ as

$$\rho_\zeta = |\zeta\rangle\langle\zeta| = \frac{1}{3}\begin{pmatrix} 1 & 1 & 0 & 1 \\ 1 & 1 & 0 & 1 \\ 0 & 0 & 0 & 0 \\ 1 & 1 & 0 & 1 \end{pmatrix}, \qquad (15)$$

where the matrices and vectors are indexed by the state labels 00, ..., 11. Now, notice that $\rho_\xi$ can be factorized as

$$\rho_\xi = \frac{1}{\sqrt{2}}\left(\begin{pmatrix} 1 & 0 \\ 0 & 0 \end{pmatrix} \otimes \begin{pmatrix} 1 & 1 \\ 1 & 1 \end{pmatrix}\right), \qquad (16)$$

where $\otimes$ is the normal tensor product. On the other hand, $\rho_\psi$ can not be factorized. States that can not be factorized are said to be entangled, while those that can be factorized are not. Notice that $\rho_\zeta$ can be partially factorized two different ways, one of which is

$$\rho_\zeta = \frac{1}{\sqrt{3}}\left(\begin{pmatrix} 1 & 1 \\ 1 & 1 \end{pmatrix} \otimes \begin{pmatrix} 0 & 0 \\ 0 & 1 \end{pmatrix} + \begin{pmatrix} 1 & 1 & 0 & 1 \\ 1 & 0 & 0 & 0 \\ 0 & 0 & 0 & 0 \\ 1 & 0 & 0 & 0 \end{pmatrix}\right) \qquad (17)$$

(the other involves equation (12) and a different remainder); however, in both cases the factorization is not complete. Therefore, $\rho_\zeta$ is also entangled, but not to the same degree as $\rho_\psi$ (because $\rho_\zeta$ can be partially factorized but $\rho_\psi$ cannot). Thus there are different degrees of entanglement and much work has been done on better understanding and quantifying it [Joz97] [Ved97]. It is interesting to note from a computational standpoint that quantum states that are superpositions of *only* basis states that are maximally far apart in terms of Hamming distance are those states with the greatest entanglement. For example, $\rho_\psi$ is a superposition of only the states $|00\rangle$ and $|11\rangle$, which have a maximum Hamming spread, and therefore $\rho_\psi$ is maximally entangled. Finally, it should be mentioned that while interference is a quantum property that has a classical cousin, entanglement is a completely quantum phenomenon for which there is no classical analog.

## 2.6. Quantum Networks

*Quantum networks* [Deu89] are one of several theoretical models of quantum computation. Others include quantum Turing machines [Ben82], and quantum cellular automata [Grö88]. In the quantum network model, each unitary operator is modeled as a quantum logic gate that affects one, two or more qubits. Schematically, this is represented as a set of quantum "wires" entering and leaving the quantum gates, reminiscent of classical logic networks. For example, figure 1 shows a network that operates on three qubits, which are represented as lines.

By convention the logic flows from left to right. The gates are represented as boxes and labeled with the name of the operator that they represent. A dot on a quantum "wire" represents a conditional upon that qubit. Therefore, in the quantum network shown in figure 1, $\hat{A}$ represents a single qubit quantum gate, $\hat{B}$ and $\hat{C}$ represent 2-qubit quantum gates, and $\hat{D}$ represents a conditional 3-qubit gate. Suppose that $\hat{A}$ is an operator that flips the state of a qubit, $\hat{B}$ is an operator that exchanges the states of two qubits, $\hat{C}$ is an operator that flips the states of two qubits if they are equal, and $\hat{D}$ is an operator that exchanges the states of two qubits if a third qubit is in the $|1\rangle$ state. When three qubits "enter" the quantum logic network, the one labeled $q_1$ first has its state flipped; then $q_1$ and $q_2$ exchange states, $q_1$ and $q_3$ have their states flipped if they are equal, and finally $q_2$ and $q_3$ exchange states if $q_1$ is in the state $|1\rangle$. Of course, if the qubits "entering" the logic array did not exist in a superposition of states, this would be no different than a classical logic sequence. However, the qubits *do* exist in a superposition of states; therefore, these gates or operations are applied to all the states in the superposition simultaneously, resulting in what has been called *quantum parallelism*. Recall that the quantum logic gate arrays are simply a schematic way to represent the time evolution of a quantum system. They are not meant to imply that quantum computation can be physically realized in a manner similar to classical logic networks. Alternatively, the network could be represented as a product of quantum operators. Since operators are applied right to left, the network of figure 1 would be represented as the operator product $\hat{D}\hat{C}\hat{B}\hat{A}$. In what follows, both the network and the product of operators representations will be used.

## 2.7. Quantum Algorithms

The field of quantum computation is just beginning to develop and offers exciting possibilities for the field of computer science -- the most important quantum algorithms discovered to date all perform tasks for which there are no classical equivalents. For example, Deutsch's algorithm [Deu92] is designed to solve the problem of identifying whether a binary function is constant (function values are either all 1 or all 0) or balanced (the function takes an equal number of 0 and 1 values). Deutsch's algorithm accomplishes the task in order $O(n)$ time, while classical methods require $O(2^n)$ time. Simon's algorithm [Sim97] is constructed for finding the periodicity in a 2-1 binary function that is guaranteed to possess a periodic element. Here again an exponential speedup is achieved. Admittedly, both these algorithms have been designed for artificial, somewhat contrived problems. Grover's algorithm [Gro96], on the other hand, provides a method for searching an unordered quantum database in time $O(\sqrt{n})$, compared to the classical bound of $O(n)$. Here is a real-world problem for which quantum computation provides performance that is classically impossible (though the speedup is less dramatic than exponential). Finally, the most well-known and perhaps the most important quantum algorithm discovered so far is Shor's algorithm for prime factorization [Sho97]. This algorithm finds the prime factors of very large numbers in polynomial time, whereas the best known classical algorithms require exponential time. Obviously, the implications for the field of cryptography are profound. These quantum algorithms take advantage of the unique features of quantum systems to provide impressive speedup over classical approaches.

## 3. Storing and Recalling Patterns in a Quantum System

Implementation of an associative memory requires the ability to store patterns in the medium that is to act as a memory and the ability to recall those patterns at a later time. This section discusses two quantum algorithms for performing these tasks.

### 3.1. Grover's Algorithm

Lov Grover has developed an algorithm for finding one item in an unsorted database, similar to finding the name that matches a telephone number in a telephone book. Classically, if there are $N$ items in the database, this would require on average $O(N)$ queries to the database. However, Grover has shown how to do this using quantum computation with only $O(\sqrt{N})$ queries. In the quantum computational setting, finding the item in the database means measuring the system and having the system collapse with near certainty to the basis state which corresponds to the item in the database for which we are searching. The basic idea of Grover's algorithm is to invert the phase of the desired basis state and then to invert all the basis states about the average amplitude of all the states [Gro96] [Gro98]. This process produces an increase in the amplitude of the desired basis state to near unity followed by a corresponding decrease in the amplitude of the desired state back to its original magnitude. The process is cyclical with a period of $\frac{\pi}{4}\sqrt{N}$, and thus after $O(\sqrt{N})$ queries, the system may be observed in the desired state with near certainty (with probability at least $1 - \frac{1}{N}$). Interestingly this implies that the larger the database, the greater the certainty of finding the desired state [Boy96]. Of course, if even greater certainty is required, the system may be sampled $k$ times boosting the certainty of finding the desired state to $1 - \frac{1}{N^k}$. Here we present the basic ideas of the algorithm and refer the reader to [Gro96] for details. Define the following operators.

$$\hat{I}_\phi = \text{identity matrix except for } \phi\phi = -1, \tag{18}$$

which simply inverts the phase of the basis state $|\phi\rangle$ and

$$\hat{W} = \frac{1}{\sqrt{2}}\begin{bmatrix} 1 & 1 \\ 1 & -1 \end{bmatrix}, \tag{19}$$

which is often called the Walsh or Hadamard transform. This operator, when applied to a set of qubits, performs a special case of the discrete fourier transform.

Now to perform the quantum search on a database of size $N = 2^n$, where $n$ is the number of qubits, begin with the system in the $|\overline{0}\rangle$ state and apply the $\hat{W}$ operator. This initializes all the states to have the same amplitude -- $\frac{1}{\sqrt{N}}$. Next apply the $\hat{I}_\tau$ operator, where $|\tau\rangle$ is the state being sought, to invert its phase. Finally, apply the operator

$$\hat{G} = -\hat{W}\hat{I}_{\overline{0}}\hat{W} \tag{20}$$

$\frac{\pi}{4}\sqrt{N}$ times and observe the system (see figure 2). The $\hat{G}$ operator inverts all the states' amplitudes around the average amplitude of all states.

### 3.1.1. An example of Grover's algorithm

Consider a simple example for the case $N = 16$. Suppose that we are looking for the state $|0110\rangle$, or in other words, we would like our quantum system to collapse to the state $|\tau\rangle = |0110\rangle$ when observed. In order to save space, instead of writing out the entire superposition of states, a transpose vector of coefficients will be used, where the vector is indexed by the 16 basis states $|0000\rangle, \cdots, |1111\rangle$. Step 1 of the algorithm results in the state

$$|\psi\rangle = (1,0,0,0,0,0,0,0,0,0,0,0,0,0,0,0).$$

In other words, the quantum system described by $|\psi\rangle$ is composed entirely of the single basis state $|0000\rangle$. Now applying the Walsh transform in step 2 to each qubit changes the state to

$$|\psi\rangle \xrightarrow{\hat{W}} |\psi\rangle = \tfrac{1}{4}(1,1,1,1,1,1,1,1,1,1,1,1,1,1,1,1),$$

that is a superposition of all 16 basis states, each with the same amplitude. The loop of step 3 is now executed $\frac{\pi}{4}\sqrt{N} \approx 3$ times. The first time through the loop, step 4 inverts the phase of the state $|\tau\rangle = |0110\rangle$ resulting in

$$|\psi\rangle \xrightarrow{\hat{I}_\tau} |\psi\rangle = \tfrac{1}{4}(1,1,1,1,1,1,-1,1,1,1,1,1,1,1,1,1),$$

and step 5 then rotates all the basis states about the average, which in this case is $\frac{7}{32}$, so

$$|\psi\rangle \xrightarrow{\hat{G}} |\psi\rangle = \tfrac{1}{16}(3,3,3,3,3,3,11,3,3,3,3,3,3,3,3,3).$$

The second time through the loop, step 4 again rotates the phase of the desired state giving

$$|\psi\rangle \xrightarrow{\hat{I}_\tau} |\psi\rangle = \tfrac{1}{16}(3,3,3,3,3,3,-11,3,3,3,3,3,3,3,3,3),$$

and then step 5 again rotates all the basis states about the average, which now is $\frac{17}{128}$ so that

$$|\psi\rangle \xrightarrow{\hat{G}} |\psi\rangle = \tfrac{1}{64}(5,5,5,5,5,5,61,5,5,5,5,5,5,5,5,5).$$

Repeating the process a third time results in

$$|\psi\rangle \xrightarrow{\hat{I}_\tau} |\psi\rangle = \tfrac{1}{64}(5,5,5,5,5,5,-61,5,5,5,5,5,5,5,5,5).$$

for step 4 and

$$|\psi\rangle \xrightarrow{\hat{G}} |\psi\rangle = \tfrac{1}{256}(-13,-13,-13,-13,-13,-13,251,-13,-13,-13,-13,-13,-13,-13,-13,-13)$$

for step 5. Squaring the coefficients gives the probability of collapsing into the corresponding state, and in this case the chance of collapsing into the $|\tau\rangle = |0110\rangle$ basis state is $.98^2 \approx 96\%$. The chance of collapsing into one of the 15 basis states that is not the desired state is approximately $.05^2 = .25\%$ for each state. In other words, there is only a $15*.05^2 \approx 4\%$ probability of collapsing into an incorrect state. This chance of success is better than the bound $1 - \frac{1}{N}$ given above and will be even better as $N$ gets larger. For comparison, note that the chance for success after only two passes through the loop is approximately 91%, while after four passes through the loop it drops to 58%. This reveals the periodic nature of the algorithm and also demonstrates the fact that the first time that the probability for success is maximal is indeed after $\frac{\pi}{4}\sqrt{N}$ steps of the algorithm.

Figure 3 represents Grover's algorithm as a quantum network. The ellipses indicate that the three operators to the right of the wavy line are those repeated $\frac{\pi}{4}\sqrt{N}$ times. Notice that the $\hat{I}$ operators require an ancillary bit.

### 3.2. Initializing the Quantum State

[Ven98b] presents a polynomial-time quantum algorithm for constructing a quantum state over a set of qubits to represent the information in a training set. The algorithm is implemented using a polynomial number (in the length and number of patterns) of elementary operations on one, two, or three qubits. Here the necessary operators are presented briefly and the reader is referred to [Ven98b] for details.

$$\hat{S}^p = \begin{bmatrix} 1 & 0 & 0 & 0 \\ 0 & 1 & 0 & 0 \\ 0 & 0 & \sqrt{\dfrac{p-1}{p}} & \dfrac{-1}{\sqrt{p}} \\ 0 & 0 & \dfrac{1}{\sqrt{p}} & \sqrt{\dfrac{p-1}{p}} \end{bmatrix}, \qquad (21)$$

where $1 \leq p \leq m$. These operators form a set of conditional transforms that will be used to incorporate the set of patterns into a coherent quantum state. There will be a different $\hat{S}^p$ operator associated with each pattern to be stored. The interested reader may note that this definition of the $\hat{S}^p$ operator is slightly different than the original. This is because in this context, we are considering associative pattern recall rather than pattern classification and therefore have no output class *per se*. Thus the phase of the coefficients becomes unimportant in this case.

$$\hat{F}^0 = \begin{bmatrix} 0 & 1 & 0 & 0 \\ 1 & 0 & 0 & 0 \\ 0 & 0 & 1 & 0 \\ 0 & 0 & 0 & 1 \end{bmatrix} \qquad (22)$$

conditionally flips the state of the second qubit if the first qubit is in the $|0\rangle$ state; another operator, $\hat{F}^1$, conditionally flips the second qubit if the first qubit is in the $|1\rangle$ state ($\hat{F}^1$ is the same as $\hat{F}^0$ except that the off-diagonal elements occur in the bottom right quadrant rather than in the top left). These operators are referred to elsewhere as Control-NOT because a logical NOT (state flip) is performed on the second qubit depending upon (or controlled by) the state of the first qubit. A complex combination of several of these operators is used to change basis states to correspond to patterns and will be termed *FLIP* in what follows (for details see [Ven98b]).

$$\hat{A}^{00} = \begin{bmatrix} 0 & 1 & 0 & 0 & 0 & 0 & 0 & 0 \\ 1 & 0 & 0 & 0 & 0 & 0 & 0 & 0 \\ 0 & 0 & 1 & 0 & 0 & 0 & 0 & 0 \\ 0 & 0 & 0 & 1 & 0 & 0 & 0 & 0 \\ 0 & 0 & 0 & 0 & 1 & 0 & 0 & 0 \\ 0 & 0 & 0 & 0 & 0 & 1 & 0 & 0 \\ 0 & 0 & 0 & 0 & 0 & 0 & 1 & 0 \\ 0 & 0 & 0 & 0 & 0 & 0 & 0 & 1 \end{bmatrix} \qquad (23)$$

conditionally flips the state of the third qubit if and only if the first two are in the state $|00\rangle$. Note that this operator is really just a Fredkin gate [Fre82] and can be thought of as performing a logical AND of the negation of the first two bits, writing a 1 in the third if and only if the first two are both 0. Three other operators, $\hat{A}^{01}$, $\hat{A}^{10}$ and $\hat{A}^{11}$, are variations of $\hat{A}^{00}$ in which the off diagonal elements occur in the other three possible locations along the main diagonal. $\hat{A}^{01}$ can be thought of as performing a logical AND of the first bit and the negation of the second, and so forth. A combination of these operators is used to identify specific states in a superposition and along with one $\hat{F}^1$ operator combine to form complex operator that will be called *SAVE* here (again for details see [Ven98b]).

Now given a set $\wp$ of $m$ binary patterns of length $n$ to be memorized, the quantum algorithm for storing the patterns requires a set of $2n+1$ qubits, the first $n$ of which actually store the patterns and can be thought of analogously as $n$ neurons in a quantum associative memory. For convenience, the qubits are arranged in three quantum registers labeled $x$, $g$, and $c$, and the quantum state of all three registers together is represented in the Dirac notation as $|x,g,c\rangle$. The algorithm proceeds as follows (see figure 4).

The $x$ register will hold a superposition of the patterns. There is one qubit in the register for each bit in the patterns to be stored, and therefore any possible pattern can be represented. The $g$ register is a garbage register used only in identifying a particular state. It is restored to the state $|\overline{0}\rangle$ after every iteration. The $c$ register contains two control qubits that indicate the status of each state at any given time and may also be restored to the $|\overline{0}\rangle$ state at the end of the algorithm. A high-level intuitive description of the algorithm is as follows. The system is initialized to the single basis state $|\overline{0}\rangle$. The qubits in the $x$ register are selectively flipped so that their states correspond to the inputs of the first pattern. Then, the state in the superposition representing the pattern is "broken" into two "pieces" -- one "larger" and one "smaller" and the status of the smaller one is made permanent in the $c$ register. Next, the $x$ register of the larger piece is selectively flipped again to match the input of the second pattern, and the process is repeated for each pattern. When all the patterns have been "broken" off of the large "piece", then all that is left is a collection of small pieces, all the same size, that represent the patterns to be stored; in other words, a coherent superposition of states is created that corresponds to the patterns, where the amplitudes of the states in the superposition are all equal. The algorithm requires $O(mn)$ steps to encode the patterns as a quantum superposition over $n$ quantum neurons. Note that this is optimal in the sense that just reading each instance once cannot be done any faster than $O(mn)$.

### 3.2.1. An example of storing patterns in a quantum system

A concrete example for a set of binary patterns of length 2 will help clarify much of the preceding discussion. Suppose that we are given the pattern set $\wp = \{01,10,11\}$. Recall that the $x$ register is the important one that corresponds to the various patterns, that the $g$ register is used as a temporary workspace to mark certain states and that the $c$ register is a control register that is used to determine which states are affected by a particular operator. Now the initial state $|00,0,00\rangle$ is generated and the algorithm evolves the quantum state through the series of unitary operations described in figure 4.

First, for any state whose $c_2$ qubit is in the state $|0\rangle$, the qubits in the $x$ register corresponding to non-zero bits in the first pattern have their states flipped (in this case only the second $x$ qubit's state is flipped) and then the $c_1$ qubit's state is flipped if the $c_2$ qubit's state is $|0\rangle$. This flipping of the $c_1$ qubit's state marks this state for being operated upon by an $\hat{S}^p$ operator in the next step. So far, there is only one state, the initial one, in the superposition, so things are pretty simple. This flipping is accomplished with the *FLIP* operator of line 3 in figure 4.

$$|00,0,00\rangle \xrightarrow{FLIP} |01,0,10\rangle$$

Next, any state in the superposition with the $c$ register in the state $|10\rangle$ (and there will always be only one such state at this step) is operated upon by the appropriate $\hat{S}^p$ operator (with $p$ equal to the number of patterns including the current one yet to be processed, in this case 3). This essentially "carves off" a small piece and creates a new state in the superposition. This operation corresponds to line 4 of figure 4.

$$\xrightarrow{\hat{S}^3} \tfrac{1}{\sqrt{3}}|01,0,11\rangle + \sqrt{\tfrac{2}{3}}|01,0,10\rangle$$

Next, the two states affected by the $\hat{S}^p$ operator are processed by the *SAVE* operator of line 5 of the algorithm. This makes the state with the smaller coefficient a permanent representation of the pattern being processed and resets the other to generate a new state for the next pattern. At this point one pass through the loop of line 2 of the algorithm has been performed.

$$\xrightarrow{SAVE} \frac{1}{\sqrt{3}}|01,0,01\rangle + \sqrt{\frac{2}{3}}|01,0,00\rangle$$

Now, the entire process is repeated for the second pattern. Again, the *x* register of the appropriate state (that state whose $c_2$ qubit is in the $|0\rangle$ state) is selectively flipped to match the new pattern. Notice that this time the generator state has its *x* register in a state corresponding to the pattern that was just processed. Therefore, the selective qubit state flipping occurs for those qubits that correspond to bits in which the first and second patterns differ -- both in this case.

$$\xrightarrow{FLIP} \frac{1}{\sqrt{3}}|01,0,01\rangle + \sqrt{\frac{2}{3}}|10,0,10\rangle$$

Next, another $\hat{S}^p$ operator is applied to generate a representative state for the new pattern.

$$\xrightarrow{\hat{S}^2} \frac{1}{\sqrt{3}}|01,0,01\rangle + \frac{1}{\sqrt{2}}\sqrt{\frac{2}{3}}|10,0,11\rangle + \sqrt{\frac{1}{2}}\sqrt{\frac{2}{3}}|10,0,10\rangle$$

Again, the two states just affected by the $\hat{S}^p$ operator are operated on by the *SAVE* operator, the one being made permanent and the other being reset to generate a new state for the next pattern.

$$\xrightarrow{SAVE} \frac{1}{\sqrt{3}}|01,0,01\rangle + \frac{1}{\sqrt{3}}|10,0,01\rangle + \sqrt{\frac{1}{3}}|10,0,00\rangle$$

Finally, the third pattern is considered and the process is repeated a third time. The *x* register of the generator state is again selectively flipped. This time, only those qubits corresponding to bits that differ in the second and third patterns are flipped, in this case just qubit $x_2$.

$$\xrightarrow{FLIP} \frac{1}{\sqrt{3}}|01,0,01\rangle + \frac{1}{\sqrt{3}}|10,0,01\rangle + \sqrt{\frac{1}{3}}|11,0,10\rangle$$

Again a new state is generated to represent this third pattern.

$$\xrightarrow{\hat{S}^1} \frac{1}{\sqrt{3}}|01,0,01\rangle + \frac{1}{\sqrt{3}}|10,0,01\rangle + \frac{1}{\sqrt{1}}\sqrt{\frac{1}{3}}|11,0,11\rangle + \sqrt{\frac{0}{1}}\sqrt{\frac{1}{3}}|11,0,10\rangle$$

Finally, proceed once again with the *SAVE* operation.

$$\xrightarrow{SAVE} \frac{1}{\sqrt{3}}|01,0,01\rangle + \frac{1}{\sqrt{3}}|10,0,01\rangle + \frac{1}{\sqrt{3}}|11,0,01\rangle$$

At this point, notice that the states of the *g* and *c* registers for all the states in the superposition are the same. This means that these registers are in no way entangled with the *x* register, and therefore since they are no longer needed they may be ignored without affecting the outcome of further operations on the *x* register. Thus, the simplified representation of the quantum state of the system is

$$-\frac{1}{\sqrt{3}}|01\rangle + \frac{1}{\sqrt{3}}|10\rangle - \frac{1}{\sqrt{3}}|11\rangle,$$

and it may be seen that the set of patterns $\mathcal{P}$ is now represented as a quantum superposition in the *x* register.

The quantum network representation of the algorithm for a single pattern is shown in figure 5. The *FLIP* operator is composed of the $\hat{F}^0$ operators left of the $\hat{S}^p$ and the question marks signify that the operator is applied only if the qubit's state differs from the value of the corresponding bit in the pattern being processed. The *SAVE* operator is composed of the $\hat{A}$ operators and the $\hat{F}^1$ to the right of $\hat{S}^p$. The network shown is simply repeated for additional patterns.

### 3.3. Grover's Algorithm Revisited

Grover's original algorithm only applies to the case where all basis states are represented in the superposition equally to start with and one and only one basis state is to be recovered. In other words, strictly speaking, the original algorithm would only apply to the case when the set $\mathcal{P}$ of patterns to be memorized includes all possible patterns of length *n* and when we know all *n* bits of the pattern to be recalled -- not a very useful associative memory. However, several other papers have since generalized Grover's original algorithm and improved on his analysis to include cases where not all possible patterns are represented and where more than one target state is to be found [Boy96] [Bir98] [Gro98]. Strictly speaking it is these more general results which allow us to create a useful QuAM that will associatively recall patterns.

In particular, [Bir98] is useful as it provides bounds for the case of using Grover's algorithm for the case of arbitrary initial amplitude distributions (whereas Grover originally assumed a uniform distribution). It turns out that a high probability for success using Grover's original algorithm depends upon this assumption of initial uniformity as the following modified version of example 3.1.1 will show.

### 3.3.1. Grover example revisited

Recall that we are looking for the state $|0110\rangle$, and assume that we do not perform the first two steps of the algorithm shown in figure 2 (which initialize the system to the uniform distribution) but that instead we have the initial state described by

$$|\psi\rangle = \tfrac{1}{\sqrt{6}}(1,0,0,1,0,0,1,0,0,1,0,0,1,0,0,1),$$

that is a superposition of only 6 of the possible 16 basis states. The loop of step 3 is now executed. The first time through the loop, step 4 inverts the phase of the state $|\tau\rangle = |0110\rangle$ resulting in

$$|\psi\rangle \xrightarrow{\hat{I}_\tau} |\psi\rangle = \tfrac{1}{\sqrt{6}}(1,0,0,1,0,0,-1,0,0,1,0,0,1,0,0,1),$$

and step 5 then rotates all the basis states about the average, which is $\tfrac{1}{4\sqrt{6}}$, so

$$|\psi\rangle \xrightarrow{\hat{G}} |\psi\rangle = \tfrac{1}{2\sqrt{6}}(-1,1,1,-1,1,1,3,1,1,-1,1,1,-1,1,1,-1).$$

The second time through the loop step 4 again rotates the phase of the desired state giving

$$|\psi\rangle \xrightarrow{\hat{I}_\tau} |\psi\rangle = \tfrac{1}{2\sqrt{6}}(-1,1,1,-1,1,1,-3,1,1,-1,1,1,-1,1,1,-1),$$

and then step 5 again rotates all the basis state about the average, which now is $\tfrac{1}{16\sqrt{6}}$ so that

$$|\psi\rangle \xrightarrow{\hat{G}} |\psi\rangle = \tfrac{1}{8\sqrt{6}}(5,-3,-3,5,-3,-3,13,-3,-3,5,-3,-3,5,-3,-3,5).$$

Now squaring the coefficients gives the probability of collapsing into the corresponding state. In this case, the chance of collapsing into the $|\tau\rangle = |0110\rangle$ basis state is $.66^2 \approx 44\%$. The chance of collapsing into one of the 15 basis states that is not the desired state is approximately 56%. This chance of success is much worse than that seen in example 3.1.1, and the reason for this is that there are now two types of undesirable states: those that existed in the superposition to start with but that are not the state we are looking for and those that were not in the original superposition but were introduced into the superposition by the $\hat{G}$ operator. The problem comes from the fact that these two types of undesirable states acquire opposite phases and thus to some extent cancel each other out. Therefore, during the rotation about average performed by the $\hat{G}$ operator the average is smaller than it should be if it were to just represent the states in the original superposition. As a result, the desired state is rotated about a suboptimal average and never gets as large a probability associated with it as it should. In [Bir98], Biron, et. al. give an analytic expression for the maximum possible probability using Grover's algorithm on an arbitrary starting distribution.

$$P_{max} = 1 - \sum_{j=r+1}^{N} |l_j - \bar{l}|^2, \tag{24}$$

where $N$ is the total number of basis states, $r$ is the number of desired states (looking for more than one state is another extension to the original algorithm), $l_j$ is the initial amplitude of state $j$, and they assume without loss of generality that the desired states are number 1 to $r$ and the other states are numbered $r+1$ to $N$. $\bar{l}$ is the average amplitude of all the undesired states, and therefore the second term of equation (24) is proportional to the variance in the amplitudes. Obviously, in the uniform case that the original algorithm assumed, the variance will be 0 and therefore $P_{max} = 1$; and in example 3.1.1 we do get 96% probability of success. The reason we do not reach the theoretical maximum is that equation (24) is a tight bound only in the case of non-integer time steps. Since this is not realistic, it becomes in practice an upper bound. Now consider the case of the initial distribution of example 3.3.1. The variance is proportional to $10*.13^2+5*.28^2 = .56$ and thus $P_{max} = .44$.

In order to rectify this problem, we modify Grover's algorithm as in figure 6. The difference between this and Grover's original algorithm is first, we do not begin with the state $|\bar{0}\rangle$ and transform it into the uniform distribution. Instead we assume some other initial distribution (such as would be the result of the pattern storage algorithm described in section 3.2). This modification is actually suggested in [Bir98]. The second modification, which has not been suggested before, is that of step 3 in figure 6. That is, the second state rotation operator not only rotates the phase of the desired states, but also rotates the phases of all the stored pattern states as well. The reason for this is to force the two different kinds of nondesired states to have the same phase, rather than opposite phases as in the original algorithm. After step 4 in figure 6, then, we can consider the state of the system as the input into the normal loop of Grover's algorithm.

With this modification of the algorithm, we can once again rework the example of 3.1.1, again starting with the state

$$|\psi\rangle = \tfrac{1}{\sqrt{6}}(1,0,0,1,0,0,1,0,0,1,0,0,1,0,0,1).$$

The first two steps are identical to those above:
$$|\psi\rangle \xrightarrow{\hat{I}_\tau} |\psi\rangle = \frac{1}{\sqrt{6}}(1,0,0,1,0,0,-1,0,0,1,0,0,1,0,0,1)$$

and
$$|\psi\rangle \xrightarrow{\hat{G}} |\psi\rangle = \frac{1}{2\sqrt{6}}(-1,1,1,-1,1,1,3,1,1,-1,1,1,-1,1,1,-1).$$

Now, all the states present in the original superposition are phase rotated and then all states are again rotated about average:
$$|\psi\rangle \xrightarrow{\hat{I}_\wp} |\psi\rangle = \frac{1}{2\sqrt{6}}(1,1,1,1,1,1,-3,1,1,1,1,1,1,1,1,1)$$

and
$$|\psi\rangle \xrightarrow{\hat{G}} |\psi\rangle = \frac{1}{4\sqrt{6}}(1,1,1,1,1,1,9,1,1,1,1,1,1,1,1,1).$$

Finally, we enter the loop of line 5 and have
$$|\psi\rangle \xrightarrow{\hat{I}_\tau} |\psi\rangle = \frac{1}{4\sqrt{6}}(1,1,1,1,1,1,-9,1,1,1,1,1,1,1,1,1)$$

for step 6 and
$$|\psi\rangle \xrightarrow{\hat{G}} |\psi\rangle = \frac{1}{16\sqrt{6}}(-1,-1,-1,-1,-1,-1,39,-1,-1,-1,-1,-1,-1,-1,-1,-1)$$

for step 7. Squaring the coefficients gives the probability of collapsing into the desired $|\tau\rangle = |0110\rangle$ basis state as 99% -- a significant improvement that is critical for the QuAM proposed in the next section.

## 4. Quantum Associative Memory

A quantum associative memory (QuAM) can now be constructed from the two algorithms of section 3. Define $\hat{P}$ as an operator that implements the algorithm of figure 4 for memorizing patterns described in section 3.2. Then the operation of the QuAM can be described as follows. Memorizing a set of patterns is simply

$$|\psi\rangle = \hat{P}|\overline{0}\rangle, \qquad (25)$$

with $|\psi\rangle$ being a quantum superposition of basis states, one for each pattern. Now, suppose we know $n$-1 bits of a pattern and wish to recall the entire pattern. We can use the modified Grover's algorithm to recall the pattern as

$$|\psi\rangle = \hat{G}\hat{I}_{\wp}\hat{G}\hat{I}_{\tau}|\psi\rangle \tag{26}$$

followed by

$$|\psi\rangle = \hat{G}\hat{I}_{\tau}|\psi\rangle \tag{27}$$

repeated $T$ times (how to calculate $T$ is covered in section 4.2), where $\tau = b_1b_2b_3?$ with $b_i$ being the value of the $i$th known bit. Since there are 2 states whose first three bits would match those of $\tau$, there will be 2 states that have their phases rotated, or marked, by the $\hat{I}_{\tau}$ operator. Thus, with $2n+1$ neurons (qubits) the QuAM can store up to $N=2^n$ patterns in $O(mn)$ steps and requires $O(\sqrt{N})$ time to recall a pattern (see figure 7). This last bound is somewhat slower than desirable and may perhaps be improved with an alternative pattern recall mechanism.

### 4.1. A QuAM Example

Suppose that we have a set of patterns $\wp$ = {0000, 0011, 0110, 1001, 1100, 1111}. Then using the notation of example 3.1.1 and equation (25) a quantum state that stores the pattern set is created as

$$|\overline{0}\rangle \xrightarrow{\hat{P}} |\psi\rangle = \frac{1}{\sqrt{6}}(1,0,0,1,0,0,1,0,0,1,0,0,1,0,0,1).$$

Now suppose that we want to recall the pattern whose first three bits are 011. Then $\tau = 011?$, and applying equation (26) gives

$$|\psi\rangle \xrightarrow{\hat{I}_{\tau}} |\psi\rangle = \frac{1}{\sqrt{6}}(1,0,0,1,0,0,-1,0,0,1,0,0,1,0,0,1),$$

$$|\psi\rangle \xrightarrow{\hat{G}} |\psi\rangle = \frac{1}{2\sqrt{6}}(-1,1,1,-1,1,1,3,1,1,-1,1,1,-1,1,1,-1),$$

$$|\psi\rangle \xrightarrow{\hat{I}_{\wp}} |\psi\rangle = \frac{1}{2\sqrt{6}}(1,1,1,1,1,1,-3,-1,1,1,1,1,1,1,1,1),$$

and

$$|\psi\rangle \xrightarrow{\hat{G}} |\psi\rangle = \frac{1}{8\sqrt{6}}(1,1,1,1,1,1,17,9,1,1,1,1,1,1,1,1).$$

At this point, there is a 96.3% probability of observing the system and finding the state $|011?\rangle$. Of course there are two states that match and state $|0110\rangle$ has a 78% chance while state $|0111\rangle$ has a 22% chance. This may be resolved by a standard voting scheme and thus we have achieved our goal -- we can observe the system to see that the completion of the pattern 011 is 0110. Notice that the loop of line 10 in figure 7 is repeated $T$ times but that in this case it was never entered because $T$ happens to be zero for this example.

Figure 8 shows a high-level quantum network for the QuAM. To the left of the longer wavy line, a high-level version of the storage algorithm described in section 3.2 is represented. The ellipses indicate that this sequence is repeated for each pattern to be stored. Notice that most of the ancillary qubit lines used in the storage algorithm do not extend to the second part because they are not needed. In other words the QuAM exists only in the $x$ register. The operators to the right of the longer wavy line show a high-level implementation of the modified Grover's algorithm described in section 3.3. The second pair of ellipses represent the fact that the final two operators will be repeated a number of times (as determined by equation (38) below).

Using some concrete numbers, assume that $n = 2^4$ and $m = 2^{14}$ (we let $m$ be less than the maximum possible $2^{16}$ to allow for some generalization and to avoid the contradictory patterns that would otherwise result). Then the QuAM requires $O(mn) = O(2^{18}) < 10^6$ operations to memorize the patterns and $O(\sqrt{N}) = O(\sqrt{2^{16}}) < 10^3$ operators to recall a pattern. For comparison, in [Bar96] Barenco gives estimates of how many operations might be performed before decoherence for various possible physical implementation technologies for the qubit. These estimates range from as low as $10^3$ (electrons in GaAs and electron quantum dots) to as high as $10^{13}$ (trapped ions), so our estimates fall comfortably into this range, even near the low end of it. Further, the algorithm would require only $2n + 1 = 2*16+1 = 33$ qubits! For comparison, a classical Hopfield type network used as an associative memory has a saturation point around $.15n$. In other words, about $.15n$ patterns can be stored and recalled with $n$ neurons. Therefore, with $n=16$ neurons, a Hopfield network can store only $.15*16 \approx 2$ patterns. Conversely, to store $2^{14}$ patterns would require that the patterns be close to 110,000 bits long and that the network have that same number of neurons. So the QuAM provides significant advantages over a classical associative memory.

The QuAM also compares favorably with other quantum computational algorithms because it requires far fewer qubits to perform significant computation that appears to be impossible classically. For example, Shor's algorithm requires hundreds or thousands of qubits to perform a factorization that can not be done classically. Vedral, et. al. give estimates for number of qubits needed for modular exponentiation, which dominates Shor's algorithm, anywhere from $7n+1$ down to $4n+3$ [Ved96]. For a 512 bit number (which RSA actually claims may not be large enough to be safe anymore, even classically), this translates into anywhere from 3585 down to 2051 qubits. As for elementary operations, they claim $O(n^3)$, which in this case would be $O(10^8)$. Therefore, the algorithm presented here requires orders of magnitude fewer operations and qubits than Shor's in order to perform significant computational tasks. This is an important result since quantum computational technology is still immature -- and maintaining and manipulating the coherent superposition of a quantum system of 30 or so qubits should be attainable sooner than doing so for a system of 2000 qubits.

As mentioned in the introduction, very recently Jones and Mosca have succeeded in physically implementing Deutsch's algorithm on a nuclear magnetic resonance (NMR) quantum computer based on the pyrimidine base cytosine [Jon98]. Even more pertinent to this work, Chuang et. al. have succeeded in physically implementing Grover's algorithm for the case $n=2$ using NMR technology on a solution of chloroform molecules [Chu98]. It is therefore not unreasonable to assume that a quantum associative memory may be implemented in the not too distant future.

### 4.2. Probability of Success

Let $N$ be the total number of basis states, $r_1$ be the number of marked states that correspond to stored patterns, $r_0$ be the number of marked states that do not correspond to stored patterns, and $p$ be the number of patterns stored in the QuAM. We would like to find the average amplitude $\bar{k}$ of the marked states and the average amplitude $\bar{l}$ of the unmarked states after applying equation (26). It can be shown that

$$k_0 = 4a - ab, \tag{28}$$

$$k_1 = 4a - ab + 1, \tag{29}$$

$$l_0 = 2a - ab, \tag{30}$$

and

$$l_1 = 4a - ab - 1. \tag{31}$$

Here $k_0$ is the amplitude of the spurious marked states, $k_1$ is the amplitude of the marked states that correspond to stored patterns, $l_0$ is the amplitude of the spurious unmarked states, $l_1$ is the amplitude of the unmarked states that correspond to stored patterns after applying equation (26), and

$$a = \frac{2(p - 2r_1)}{N}, \tag{32}$$

and

$$b = \frac{4(p + r_0)}{N}. \tag{33}$$

A little more algebra gives the averages as

$$\bar{k} = 4a - ab + \frac{r_1}{r_0 + r_1}, \tag{34}$$

and

$$\bar{l} = -ab + \frac{2a(N + p - r_0 - 2r_1)}{N - r_0 - r_1} - \frac{(p - r_1)}{N - r_0 - r_1}. \tag{35}$$

Now we can consider this new state described by equations (28-31) as the arbitrary initial distribution to which the results of [Bir98] can be applied. These can be used to calculate the upper bound on the accuracy of the QuAM as well as the appropriate number of times to apply equation (27) in order to be as close to that upper bound as possible. The upper bound on accuracy is given by

$$P_{\max} = 1 - (N - p - r_0)|l_0 - \bar{l}|^2 - (p - r_1)|l_1 - \bar{l}|^2, \tag{36}$$

whereas the actual probability at a given time $t$ is

$$P(t) = P_{\max} - (N - r_0 - r_1)|\bar{l}(t)|^2. \tag{37}$$

The first integer time step $T$ for which the actual probability will be closest to this upper bound is given by rounding the function

$$T = \frac{\frac{\pi}{2} - \arctan\left(\frac{\bar{k}}{\bar{l}}\sqrt{\frac{r_0 + r_1}{N - r_0 - r_1}}\right)}{\arccos\left(1 - 2\frac{r_0 + r_1}{N}\right)} \tag{38}$$

to the nearest integer.

### 4.3. Extension to Non-binary Patterns

The algorithm described in section 4.1 can handle only binary patterns. Nominal data with more than two values can be handled one of two ways: convert the multiple values into a binary representation or extend the algorithm to handle data of more than two values. In [Ven98b] a generalization of the pattern storage algorithm to more that two values is presented; if Grover's algorithm can likewise be generalized then a straightforward generalization of the QuAM to the case of nonbinary patterns can easily be conceived. This may be preferable to converting to a binary format since doing so would introduce a number of new qubits, requiring larger quantum registers which could result in degradation of pattern completion accuracy.

### 5. Concluding Comments

A unique view of the associative pattern completion problem is presented that allows the proposal of a quantum associative memory with exponential storage capacity. It employs simple spin-1/2 (two-state) quantum systems and represents patterns as quantum operators. This approach introduces a promising new field to which quantum computation may be applied to advantage -- that of neural networks. In fact, it is the authors' opinion that this application of quantum computation will, in general, demonstrate greater returns than its application to more traditional computational tasks (though Shor's algorithm is an obvious exception). We make this conjecture because results in both quantum computation and neural networks are by nature probabilistic and inexact, whereas most traditional computational tasks require precise and deterministic outcomes.

This paper presents a quantum computational learning algorithm that takes advantage of the unique capabilities of quantum computation to produce an important advance in the field of neural networks. In other words, the paper makes an important contribution to both the field of neural computation and to the field of quantum computation -- producing both a new neural network result and a new quantum algorithm that accomplishes something that no classical algorithm has been able to do -- creating a reliable associative memory with a capacity exponential in the length of the patterns to be stored.

The most urgently appealing future work suggested by the result of this paper is, of course, the physical implementation of the algorithm in a real quantum system. As mentioned in sections 1 and 4, the fact that very few qubits are required for non-trivial problems together with the recent physical realization of Grover's algorithm helps expedite the realization of quantum computers performing useful computation. In the mean time, as discussed in section 4, the time bound for recall of patterns is slower than desirable, and alternatives to Grover's algorithm for recalling the patterns are being investigated. Also, a simulation of the quantum associative memory may be developed to run on a classical computer at the cost of an exponential slowdown in the length of the patterns. Thus, association problems that are non-trivial and yet small in size will provide interesting study in simulation. We are also investigating associative recall of nonbinary patterns using spin systems higher than 1/2 (systems with more than two states). Another important area for future research is investigating further the application of quantum computational ideas to the field of neural networks -- the discovery of other quantum computational learning algorithms. Further, techniques and ideas that result from developing quantum algorithms may be useful in the development of new classical algorithms. Finally, the process of understanding and developing a theory of quantum computation provides insight and contributes to a furthering of our understanding and development of a general theory of computation.

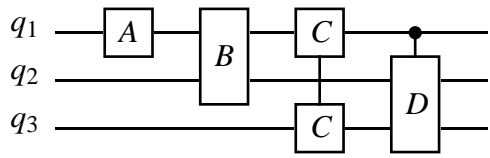

1. Generate the initial state $|\bar{0}\rangle$
2. $\hat{W}|\bar{0}\rangle = |\bar{1}\rangle = |\psi\rangle$
3. Repeat $\frac{\pi}{4}\sqrt{N}$ times
4. $\quad\quad |\psi\rangle = \hat{I}_\tau |\psi\rangle$
5. $\quad\quad |\psi\rangle = \hat{G}|\psi\rangle$
6. Observe the system

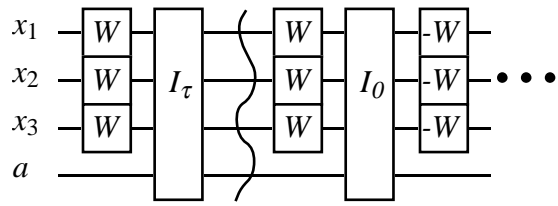

1. Generate the initial state $|\psi\rangle = |\bar{0}\rangle$
2. For $m \geq p \geq 1$
3.     $FLIP|\psi\rangle$
4.     $\hat{S}^p|\psi\rangle$
5.     $SAVE|\psi\rangle$

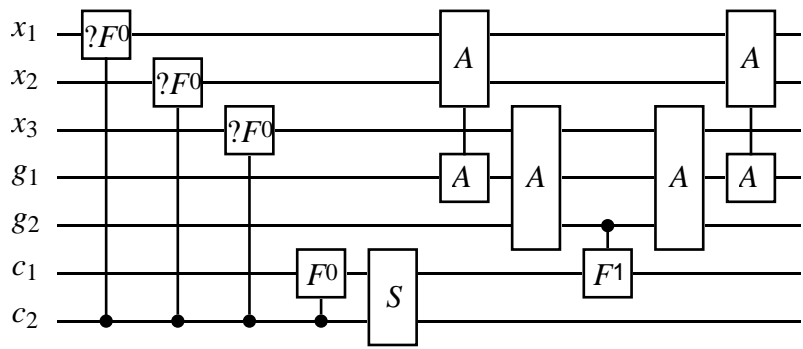

1. $|\psi\rangle = \hat{I}_\tau |\psi\rangle$
2. $|\psi\rangle = \hat{G} |\psi\rangle$
3. $|\psi\rangle = \hat{I}_\wp |\psi\rangle$
4. $|\psi\rangle = \hat{G} |\psi\rangle$
5. Repeat $\frac{\pi}{4}\sqrt{N}$ -2 times
6. $|\psi\rangle = \hat{I}_\tau |\psi\rangle$
7. $|\psi\rangle = \hat{G} |\psi\rangle$
8. Observe the system

1. Generate the initial state $|\psi\rangle = |\bar{0}\rangle$
2. For $m \geq p \geq 1$
3.     $FLIP|\psi\rangle$
4.     $\hat{S}^p|\psi\rangle$
5.     $SAVE|\psi\rangle$
6. $|\psi\rangle = \hat{I}_\tau|\psi\rangle$
7. $|\psi\rangle = \hat{G}|\psi\rangle$
8. $|\psi\rangle = \hat{I}_\wp|\psi\rangle$
9. $|\psi\rangle = \hat{G}|\psi\rangle$
10. Repeat $T$ times
11.     $|\psi\rangle = \hat{I}_\tau|\psi\rangle$
12.     $|\psi\rangle = \hat{G}|\psi\rangle$
13. Observe the system

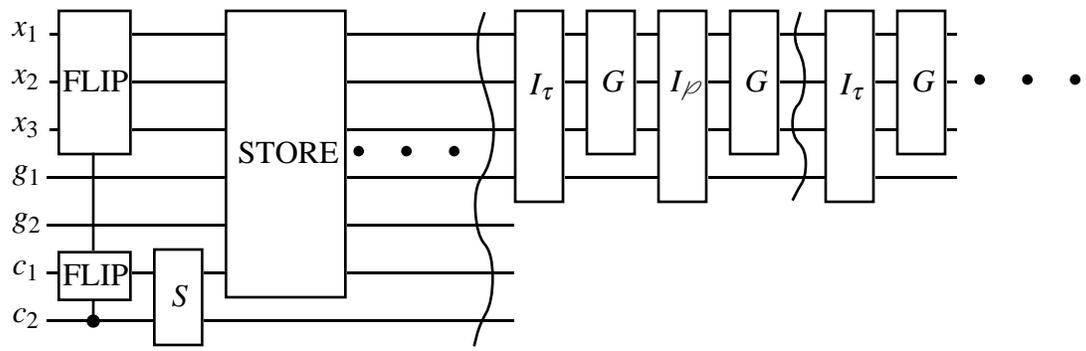